\documentclass[twocolumn]{revtex4}

\usepackage{amsmath}    
\usepackage{graphicx}   
\usepackage{verbatim}   
\usepackage{color}      
\usepackage{subfigure}  
 
\def\lowsim{\mathrel{\lower 0.7 ex \hbox to 0 pt{$\sim$\hss}}}

 \newcommand{\wpi}{\omega_{pi}}

 \newcommand{\m}[1]{\mathbf #1} 

\begin{document}

\title{Propinquity of current and vortex structures: 
effects on collisionless plasma heating}
\author{Tulasi N Parashar}
\author{William H Matthaeus}
\affiliation{Bartol Research Institute, Department of Physics and Astronomy, University of Delaware, Newark, DE}

\begin{abstract}
Intermittency of heating in weakly collisional plasma turbulence
is an active subject of research, with significant potential impact
on understanding of the solar wind, solar corona and astrophysical
plasmas.  Recent studies suggest a role of vorticity in plasma heating.
In magnetohydrodynamics small scale vorticity is generated near current
sheets and this effect persists in kinetic plasma, as demonstrated
here with hybrid and fully kinetic Particle-In-Cell (PIC) simulations.
Furthermore, vorticity enhances local kinetic effects, with a generalized
resonance condition selecting sign-dependent enhancements or reductions
of proton heating and thermal anisotropy.  In such plasmas heating
is correlated with vorticity and current density, but more strongly
with vorticity.  These results help explain several prior results that
find kinetic effects and energization near to, but not centered on,
current sheets.  Evidently intermittency  in kinetic plasma involves
multiple physical quantities, and the associated coherent structures
and nonthermal effects are closely related.
\end{abstract}

\maketitle
\section{Introduction}
Weakly collisional plasmas in planetary magnetoshperes
\citep[e.g.][]{GaleevPHPM75, SaurAA02, RetinoNature07},
stellar coronae \citep[e.g.][]{ShimizuPASJ95}, solar and
stellar winds \citep[e.g.][]{TuSSR95}, black hole accretion
disks \citep[e.g.][]{SharmaApJ03}, galactic cooling flows
\citep[e.g.][]{CarilliARAA02}, and elsewhere, are observed
\citep[e.g.][]{SaurAA02, RetinoNature07, ColemanApJ68, BrunoLRSP13,
MarschLRSP06} or believed to be \citep[e.g.][]{MacLowApJ99,
BanerjeeMNRAS14} in a turbulent state.  These systems are typically
more heated than what is expected from simplistic considerations
\citep[e.g.][]{BarnesApJ68, BarnesApJ69, HellingerJGR11, SharmaMNRAS12},
a property consistent with turbulent cascade and dissipation.
In these plasmas the conversion of larger fluid scale energy
into thermal degrees of freedom must be accomplished by particle
interactions with electromagnetic fields, leading ultimately to
thermalization by collective effects or infrequent collisions
\citep[e.g.][]{KlimontovichPU97,SchekochihinApJS09,HowesPTRSA15}.
Understanding how collisionless processes lead to dissipation is
an active research topic that is traditionally studied using linear
Vlasov theory.  A recent emphasis has been the study of intermittent,
or spatially nonuniform, processes that occur in non-volume filling
coherent structures, in analogy to intermittency properties in
hydrodynamics.  A specific focus of this work has been on electric
current structures such as current filaments or current sheets.
Intense kinetic activity, including temperature anisotropy, is found {\it
near}  current enhancements (sheets), but not necesssarily centered on
them \cite{ServidioPRL12, GrecoPRE12, ServidioApJL14}.  Studies have also
identified plasma flow vorticity and shear as important elements in
heating and generation of kinetic effects \cite{HubaGRL96,MarkovskiiApJ06,
DelSartoPRE16,FranciSW14}.  Here we show that  kinetic plasma generates
vorticity structures, typically very close to current sheets or filaments.
We will demonstrate that protons are preferentially heated in one sign
of vorticity, and cooled in the other, confirming that the sign of
$\omega \cdot {\bf B}$ controls this effect \cite{HubaGRL96,DelSartoPRE16}.
This provides a unified interpretation of a number of recent results in
the theory and simulation of kinetic plasma turbulence.

\section{Background}
Various processes may contribute to the collisionless transfer of
energy from fields into particles, e.g., wave-particle interactions
\citep[e.g.][]{HollwegJGR02, HowesPTRSA15} \citep[]{StixBook}, heating
at intermittent structures \citep[e.g.][]{ParasharSW12, ParasharPP11,
OsmanApJL11, WanPRL15}, stochastic orbits \citep{ChenJGR86},
stochastic energization by electromagnetic fields at kinetic scales
\citep[e.g.][]{ChandranApJ10-1, BourouaineApJ13}.  While it is not
clear whether a single process is dominant under varying conditions
(e.g. plasma $\beta$, Mach number, Alfv\'en ratio, etc.), a broader
question is whether heating is mainly due to distributed 
homogeneous processes
depending on average plasma properties (e.g., from linear Vlasov theory),
or if the main contributors are inherently inhomogeneous localized 
processes,
depending on gradients and intermittency of the turbulence.

Recent years have seen increasing evidence for
intermittent heating, gathered from observations
\citep[e.g.][]{SundkvistPRL07,OsmanApJL11,OsmanPRL12,WuApJL13} as well
as several types of numerical simulations \citep[e.g.][]{ParasharSW12,
ParasharPP11, ServidioPRL12, GrecoPRE12,TenBargeApJL13,WanPRL12,
KarimabadiPP13,ServidioApJL14,WanPRL15}.  Efforts to {\em quantify}
the nature of intermittency have also been made using MHD as well
as kinetic simulations \citep[e.g.][]{ZhdankinPRL12, MakwanaPP15}.
Most of these studies concentrate on coherent magnetic structures such as
current sheets, reconnection sites, or ``PVI'' events \citep{GrecoGRL08}.
However, some studies have also examined the role of structure in the
(proton) velocity field.  Studies have examined heating of protons in
velocity shears \cite{MarkovskiiApJ06}, and the relation of enhanced
heating to vorticity \cite{FranciSW14}.  Hybrid simulations suggest
\citep{VasquezApJ12} that the kinetic heating of protons might be a
``viscous like" process instead of a magnetic process.  This is not be
unexpected since several types of coherent structures should emerge in
turbulence and kinetic effects may appear in association with any, or
all, of them, as can be readily demonstrated in solar wind observations
\citep{ServidioApJL14}.

Here we focus on vorticity, and show that vorticity structures appearing
in kinetic plasma turbulence (in conditions roughly resembling the
corona or solar wind) are intimately related to local current density
enhancements.  This effect that can be understood purely in terms of
MHD processes \citep[e.g.][]{MatthaeusGRL82}.  Plasma vortex structures
are generated very close to current structures, typically taking on a
quadrupolar configuration, and leading to plasma heating and generation
of anisotropies in temperatures and pressure.  Examining this effect
in detail will lead to better understanding of 
dynamical activity near current sheets 
\cite{ServidioPRL12,GrecoPRE12,KarimabadiPP13}
including the 
involvement of vorticity in kinetic
dissipation \cite{MarkovskiiApJ06,DelSartoPRE16,FranciSW14}.

\section{Kinetic Simulations}
We employ two types of kinetic codes, hybrid particle-in-cell (PIC) and
full PIC simulations.  Both types make use of the P3D family of codes
\cite{ZeilerJGR02}, in hybrid PIC \citep[e.g.][]{ParasharPP11} mode,
and fully kinetic PIC mode \citep[e.g.][]{WuPRL13}.  All simulations
discussed here are performed in the 2.5D geometry (2 dimensional grid
and all three components of field vectors). The hybrid simulation
has $L_x=L_y=204.8 d_i$ (where $d_i=c/\wpi$ is the ion
inertial length, with $c$ the speed of light and $\wpi$ the proton plasma
frequency), $N_x=N_y=2048$, 200 particles per cell,
$\beta_i=1.0$, cold isothermal electrons with $m_e/m_i=1/25$. The
simulation is initialized with energy only in wavevectors $\m{k}$
that have $|\m{k}| =3,4,5$. $v$ and $b$ fluctuations are chosen with
a specified initial spectral shape, gaussian random phases, and only
in essentially incompressive modes of the system.  This simulation was
also used in a recent study of variance anisotropy in kinetic plasmas
\cite{ParasharApJ16}. The first full PIC simulation has $L_x=L_y=20.48
d_i$, $N_x=N_y=1024$, 200 particles per cell, $\beta_i=\beta_e=0.08$,
$m_e/m_i=1/25$. The initial condition is Orszag-Tang vortex (OTV)
\citep[e.g.][]{OrszagJFM79,DahlburgPFB89,ParasharPP09,VasquezApJ12}. This
simulation was performed for a recent study of transition from kinetic
to MHD like behavior \cite{ParasharApJ15}. The final PIC simulation
(Turb812) has $L_x=L_y=25.6 d_i$, $N_x=N_y = 2048$, 400 particles per
cell, $\beta_i=\beta_e=0.25$, $m_e/m_i=1/25$. The latter initial condition
is MHD like, and more ``turbulent'', with $v$ and $b$ fluctuations
excited in a band of wave-vectors with $2 \le |k| \le 4$ with a specified
initial spectrum. This simulation was done as part of a recent study that
discussed the relation of time scales at proton gyroscale and their
relation to relative proton-electron heating \cite{MatthaeusApJL16}.
Particle in cell codes have an inherent noise associated with them
due to finite number of particles per cell. While performing these
simulations, the two most important numerical criteria that we
paid attention to were: i) excellent conservation of total energy
(less than a few percent of change in any fluctuation energy) and, ii)
the particle noise in the spectrum was significant only at scales
much smaller than the scales of interest (Debye length $\lambda_d$
for PIC and $d_i$ for hybrid PIC). On this basis, the modest number 
of particles employed here was considered adequate. As an additional 
measure, we employed filtering \citep[e.g.][]{WanPRL12} to remove 
particle noise at grid scales prior to computing gradients (e.g. 
vorticity).

\section{Vortex quadrupole generation } 
The most commonly discussed coherent structures in plasma physics
are current sheets and reconnection sites. However, these 
current concentrations
are also locations of strong vorticity generation
\citep[e.g.][]{MatthaeusGRL82}.  This phenomenon, often overlooked
in reconnection studies, may be completely understood in terms of two
dimensional MHD, which we briefly review here.

The relevant equations of motion for two dimensonal (2D) incompressible
MHD can be written as \citep[e.g.][]{FyfeJPP76,CarboneLNP09}:

\begin{eqnarray}
\frac{\partial \omega}{\partial t}+{\bf v}\cdot \nabla {\omega} & = & {\bf b} \cdot \nabla j_z 
+ \nu \nabla^2 \omega \label{eq:om} \\
\frac{\partial a}{\partial t}+{\bf v}\cdot \nabla {a}& = & \mu \nabla^2 a
\label{eq:a}
\end{eqnarray}
in terms of the $z$-component of vorticity $\omega = (\nabla \times
{\bf v})_z$ and the magnetic potential $a({\bf x},t)$, where the
fluctuating magnetic field is ${\bf b} = \nabla \times a({\bf x}, t)
\hat z$.  The terms involving resistivity $\mu$ and viscosity $\nu$
are dissipative and become important at small
scales.  Although highly simplified, these contain the basic physics
that establishes sites of magnetic reconnection, as in for example the
well-known Sweet Parker model \citep{ParkerBook79}.

An  important feature of the vorticity equation (\ref{eq:om}) is that when
${\bf b} \cdot \nabla j_z = 0$, Kelvin's theorem applies and vorticity
cannot be generated in the interior of the magnetofluid.  However the
term ${\bf b} \cdot \nabla j_z$, the curl of the Lorentz force, can
generate vorticity, just as sheared velocity fields can stretch magnetic
field lines and amplify the mean square current density.  Taken together
these two effects are responsible for amplification of gradients and the
enhancement of dissipation that drive important relaxation processes in
MHD \cite{TaylorPRL74, MatthaeusANYAS80,BiskampBook03}.

\begin{figure}
\begin{center}
\includegraphics[width=\columnwidth]{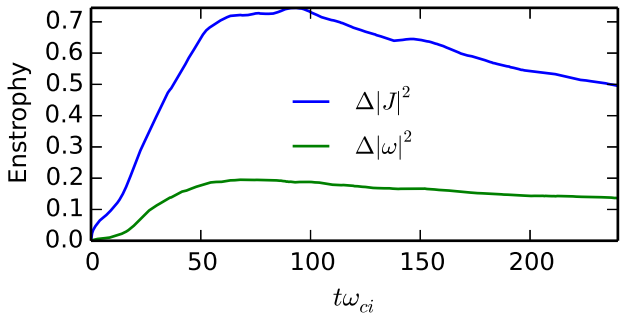}
\caption{Current grows before vorticity.}
\label{enst}
\end{center}
\end{figure}

Figure (\ref{enst}) provides a simple demonstration that vorticity
generation is also operative in a kinetic plasma.  The Figure shows
the change in magnetic enstrophy (mean square electric current density
$\Omega_b = \langle |J|^2 \rangle$ ) and the enstrophy (mean square
vorticity $\Omega_v = \langle |\omega|^2 \rangle $) measured relative
to their respective initial values for the hybrid run.  Both $\Omega_b$
and $\Omega_v$ increase until a maximum value is reached after one or
two nonlinear times. This behavior is familiar from MHD turbulence
simulations that begin from band limited initial conditions (e.g.,
\cite{MatthaeusANYAS80,BiskampBook03}), the peak time being called in that
case ``the peak of dissipation''.  Note that prior to the peak time, the
mean square current increases somewhat more rapidly than the enstrophy.
This behavior is common to all the kinetic turbulence simulations of this
type, and also in MHD turbulence simulations with band limited initial
data and fairly large and approximately equal Reynolds and magnetic
Reynolds numbers.  In both MHD and the kinetic case, this effect is
indicative of the vorticity generation that is secondary to current
sheet formation.

\begin{figure}[!hbt]
\begin{center}
\includegraphics[width=\columnwidth]{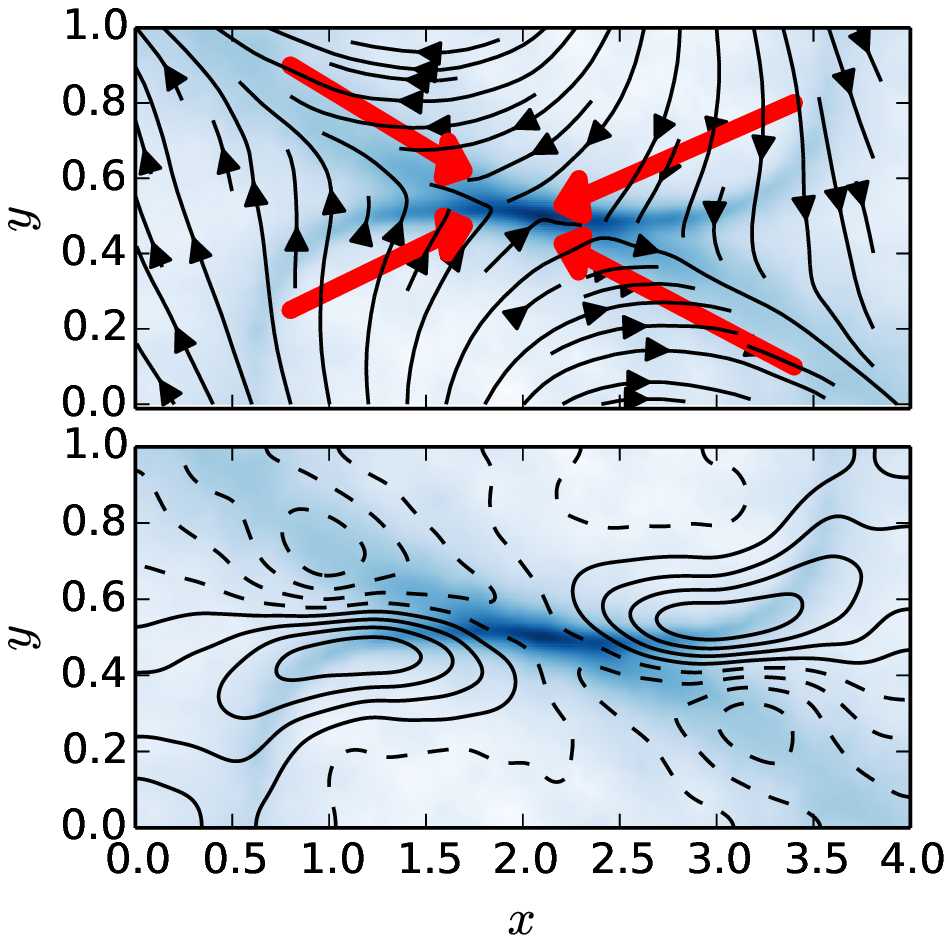}
{Central current sheet from PIC OTV simulation (rotated by 30 degrees for clarity). $x,y$ coordinates in arbitrary units. (Top) Magnetic field lines (black); and approximate directions of $\nabla j_z$ (red arrows). (Bottom) Smoothed contours of vorticity (solid positive, dashed negative) superposed over shading for current intensity. ${\bf b} \cdot \nabla j_z$ matches the signs of locally generated vorticity.}
\label{vq-cartoon}
\end{center}
\end{figure}

To further discuss vorticity generation, consider a purely 2D, or
nearly 2D, geometry, and consider two oppositely directed flux tubes
that encounter each other, generating a current enhancement (say, in
the positive z-direction) with a single maximum centered on the origin
in the $x$-$y$ plane, reminiscent of typical 2D reconnection geometry.
Then $\nabla j_z$ is directed towards the origin in all four quadrants.
The main magnetic field of the tubes is such that $b_x < 0$ for $y>0$
and $b_x>0$ for $y<0$ (and we neglect the axial $z$- component of
magnetic field for this demonstration.) One recognizes immediately
\citep{MatthaeusGRL82} that the generation term ${\bf b} \cdot \nabla
j_z$ is positive in the first and fourth quadrants, and negative in
the second and fourth quadrants. That is a standard scenario of flux
tube interactions giving rise to a current enhancement (filament or
sheet). Once the current enhancement is formed, will also generate a
vorticity quadrupole structure as shown in Figure \ref{vq-cartoon}.  

It is noteworthy that the vorticity
structures produced in this way necessarily form on the flanks
of strong current enhancements. Moreover, the thickness of vortex
structures produced in this way should be comparable to the thickness
of the current sheet that drives them, in this case a few $d_i$. The
similarity of vortex and current sheet thicknesses has been noted for
example in the low beta weakly three dimensional Reduced MHD description
of turbulence \cite{WanApJ14}.  Furthermore qualitative evidence for this
can be seen in fully kinetic PIC simulations \cite{KarimabadiPP13} that
electron vorticity structures as well as current structures extend down
to electron scales, although correlations were not computed in that study.

\begin{figure*}[!hbt]
\begin{center}
\includegraphics[width=\textwidth]{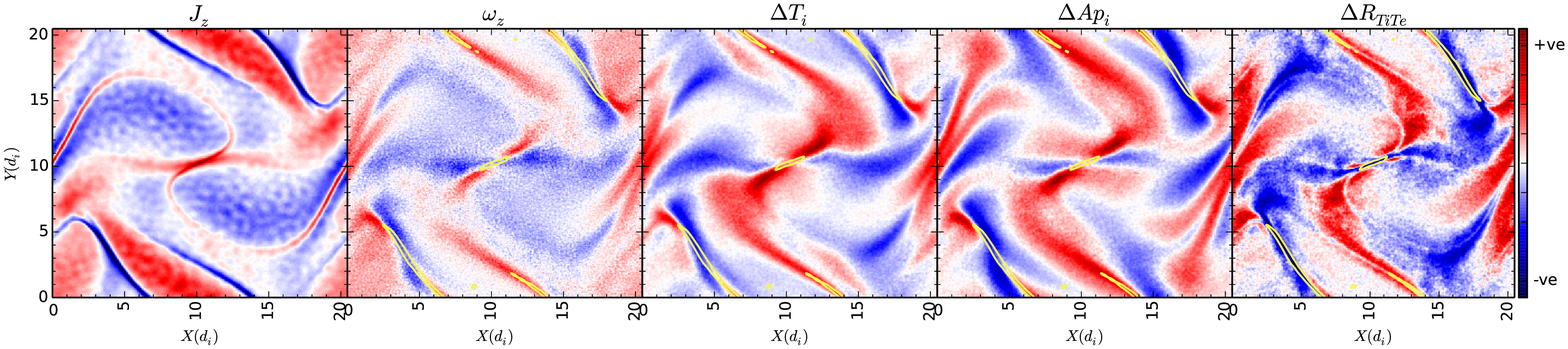}
\includegraphics[width=0.8\textwidth]{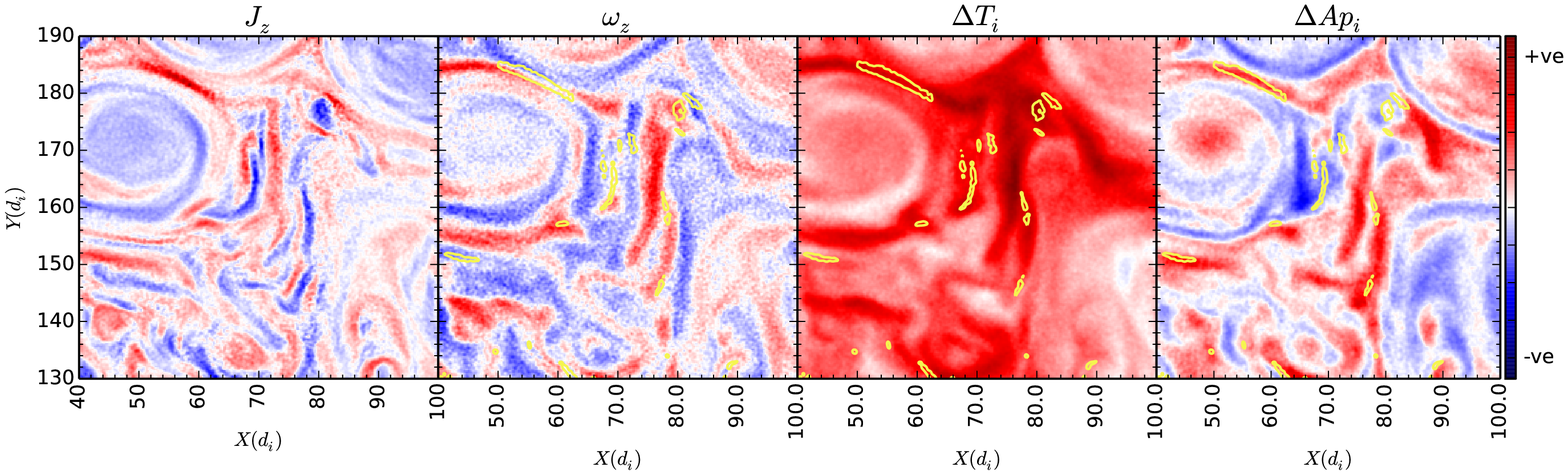}
\caption{Intermittency, or concentration into coherent structures of
various quantities, from kinetic simulations.  (Top) Orszag Tang PIC run,
an orderly configuration in which structures and their relationships are
relatively easy to identify at early times; (Left to right:) current
density $J_z$; out of plane vorticity $\omega_z$ proton temperature
increase $\Delta T_i$; change in proton temperature anisotropy $\Delta
Ap_i$, and a measure of relative proton and electron heating.  (Bottom)
small portion of a hybrid turbulence run, in which structures are
more complex and require statistical analysis. (Left to right: $J_z$,
$\omega_j$, $\Delta T_i$, and $\Delta Ap_i$.
The notation $+ve$ and $-ve$ is shorthand for 
positive and negative, respectively.)
 }
\label{jzwap}
\end{center}
\end{figure*}

The generation of vortex quadrupoles is a purely magnetohydrodynamic
phenomenon that does not require any higher order effects like
compressibility, Hall physics or any other kinetic effects.  Therefore it
is not entirely surprising that we can readily demonstrate that vortex
quadrupole formation also occurs near current sheets in kinetic plasma
simulation.  For this we refer to the first two panels on the left
side of each row in Figure (\ref{jzwap}).  The top row shows an early
time snapshot of the relatively ordered Orszag-Tang initial condition
\cite{OrszagJFM79,ParasharPP09}, here simulated with a PIC code.
The dynamics results immediately in several orderly current filaments,
for example, the one that is found in the center of the simulation (first
column, first row).  One immediately sees clean vortex quadrupoles (second
column, first row) forming near each current sheet, precisely as expected
from the above reasoning.  The second row shows small portion of a much
less orderly turbulence-style initial condition, from a hybrid PIC run.
Here one sees many current sheets and filaments, highly distorted,
and also a complex pattern of vorticity.  Although in  this case it is
difficult to recognize the pattern of vorticity generation, we will see
below that correlation statistics account for an interpretation very
similar to what is seen in the simpler Orszag-Tang case.

\section{Vorticity and expected kinetic effects}

Vortex quadrupoles form in kinetic plasma very much in the same way that
they form in MHD, on the flanks of current sheets.  Therefore one may
also expect distinctive kinetic features to appear, such as `viscous
like" heating and thermal energy transport, given that the sharpest
current sheets often form with a thickness a few times the ion inertial
scale $d_i$.  The basic mechanism for generating these kinetic effects
can be thought of as a nonlinear generalization of Landau resonance,
as one may understand based on a simple physical argument.

Protons gyrate in a left-handed sense about a uniform magnetic field.
Fluid elements, defined here as parcels of plasma following 
bulk motion
as defined by the first moment of the distribution funcion, rotate in a
right handed sense relative to the local vorticity vector $\bf \omega$.
Therefore in locations where $\m{\omega}\cdot\m{B}_0$ is positive, the
protons gyrate opposite to the vortical motion of the fluid element.  In a
low collisionality or collisionless plasma, individual particles are not
tightly coupled to fluid elements as they are in collisional gas dynamics,
so under varying circumstances, protons may change fluid elements more
or less rapidly. Clearly when the senses of rotation of particle and
fluid element are mismatched, the protons have a higher probability of
jumping from one fluid element to another one.  A given fluid element
will lose particles rapidly, but also gain particles that originate in
elements that have different average velocities.  Regionally there will
be an exchange and interchange of fluid energy and random thermal energy.
Intuitively this results either in an enhancement of heat transport,
or an increase of thermal energy, or both.

The opposite happens in locations where $\m{\omega}\cdot\m{B}_0$ is
negative and hence the vortical motion of the fluid element and the
proton gyration match.  In this case the proton distribution has a higher
probability of losing energy.  This can be understood in a simple way.
For definiteness suppose that plasma beta is unity and a current sheet
of a few $d_i$ thickness generates a (circular) vortex enhancement with
similar dimension. Then particles gyrating in the same sense of the vortex
will remain within the vortex provided that their thermal speed is not
much larger than the Alfv\'en speed.  But higher energy particles -
those in the tail of the distribution -- will rapidly scatter away as
their gyro-orbits exceed the dimension of the vortex.  Similarly cooler
particles that scatter into the vortex will easily be picked up by the
fluid due to the $E \times B$ drift  effect. The next effect is that
the fluid elements in a $\m{\omega}\cdot\m{B}_0 <0$ vortex should have
cooler protons.

Indeed, previous studies have demonstrated that there is a viscous-like 
effect associated with shear \cite{MarkovskiiApJ06,DelSartoPRE16}
or vorticity \cite{HubaGRL96} including the dependence on the sign
of ${\bf \omega} \cdot {\bf B_0}$.  A recent hybrid simulation study
\cite{FranciSW14} confirms that there is an association of kinetic
effects, including proton temperature and proton temperature anisotropy,
with this signed quantity \citep{FranciSW14}.

At this point it is useful to recall a well-known general relationship
that exists between {\it global} increase of internal energy and
velocity gradients. This can be obtained from the Vlasov equation in
suitable (e.g., periodic) boundaries using elementary manipulations.
Specifically, it is straightforward to show that the kinetic energy
$nm|{\bf v}_{th}|^2$ associated with random motions of any one species
(say, protons) obeys the equation \cite{HasegawaSatoBook}

\begin{eqnarray}
\frac {d}{dt} \frac {\langle nm|{\bf v}_{th}|^2\rangle}{2}  & = & 
 \frac {d}{dt} \frac{m}{2} \int d^3x \int d^3v  f({\bf x}, {\bf v}, t)
|{\bf v} - {\bf u}|^2 \nonumber \\
 = \langle u_i \nabla_j \Pi_{ij} \rangle 
& = &  - \langle \Pi_{ij} \nabla_i u_j \rangle,
\label{eq:random}
\end{eqnarray}
in terms of the particle velocity ${\bf v}$, the fluid velocity of this
species $\bf u$, and the pressure tensor $\Pi_{ij}$.
 The angle brackets $\langle \dots \rangle$ denote a
volume average.  Since the pressure tensor is always symmetric in indices,
the full contraction $P_{ij}\nabla_iu_j$ involves contributions only from
the symmetric part of $\nabla_i u_j$.  In general one can write $\nabla_i
u_j = C\delta_{ij} + \frac12 S_{ij} + \frac12 \sigma_{ij}$ where $C=\nabla
\cdot {\bf u}$ represents the compressional part, $S_{ij} = \nabla_i u_j +
\nabla_j u_i - 2C\delta_{ij}$ is the traceless symmetric stress tensor,
and the antisymmetric stress is $\sigma_{ij} = \epsilon _{ijk}\omega_k$.

The several considerations described in the present section and the
previous one, including the references, provide the basis for the next
section, in which we further unravel the complex kinetic processes that
occur near coherent structures in turbulence.  We should remark at this
point that our main effort here pertains to current and vortex structures,
but reconnecting current sheets will typically also occur near in regions
where one also finds shocks, compressions and various linear waves, e.g.,
in the reconnection exhausts. All these phenomena may occur at least
roughly in the vicinity of a current sheet.  The typical location of these
secondary structures is within a few $d_i$s  of a current sheet and hence practically the same location
in a system that extends to many decades in $d_i$. In large systems
(large Reynolds numbers) the sites of vorticity generation may be so
small that they are difficult to identify and hardly distinguishable
in location from their companion current sheets \citep[e.g.][]{WanApJ14}.

\section{Statistics and location of kinetic effects}

The proton temperature effects of vorticity generated near current sheets
are clearly demonstrated in the case of the orderly Orszag-Tang vortex,
as was the presence of the vortex quadrupole itself.  In the third
column, top row of Figure \ref{jzwap}, we see an illustration of the
change in proton temperature from its initial value $\Delta T_i = T_i -
T_i(t=0)$ for this case, with the analogous plot from the more complex
turbulence run in the second row.  In the Orszag-Tang case one clearly
sees that positive vorticity lobes (${\bf \omega} \cdot {\bf B}_0 >0$
of the central quadrupole are hotter, while the corresponding negative
lobes are cooler.  The fourth panel on each row of the same figure shows
differential proton anisotropy $\Delta A_{p_i} = T_{i\perp}/T_{i\parallel}
- 1$ for each case.  We also observe that the perpendicular anisotropy
is greater in the hotter $\omega >0$ lobes while the cooler $\omega<0$
lobes have parallel anisotropy (negative $\Delta A_p$).  It is of
course tempting  to associate this phenomenon with the vorticity, as
discussed by \citet[e.g.][]{FranciSW14}. We will discuss a nuance of
this interpretation further below.

An additional diagnostic in the Orszag Tang PIC run is the relative
heating of protons and electrons, shown in the final panel, top row of
Figure \ref{jzwap}. Out of many possible ways to quantify relative proton
electron heating, we choose to work with differential relative heating
defined by $\Delta R_{TiTe} = (T_iT_{e0})/(T_eT_{i0}) -1$, where $T_{i0}$
and $T_{e0}$ are the initial proton and electron temperatures. $\Delta
R_{TiTe}$ is positive where the percentage increase in proton temperature
is more than percentage change in electron temperature and vice versa. 
In
the last panel of top row in Figure \ref{jzwap} it is evident that the
${\bf \omega} \cdot {\bf B}_0 >0$ lobes, with hot, more anisotropic
protons, also have a proton population heated more than the
electron population.  The cooler, ${\bf \omega} \cdot {\bf B}_0 < 0$ lobes
have more electron heating than proton heating.  
In fact, beginning at the central current sheet (near the X-point)
and moving outward along the current sheet flanks, 
one sees first (at the center) hotter electrons, then a region 
with almost equal proton and electron heating (along the flanks), and 
finally (outside the flanks) regions of hotter protons.  
Clearly, the organization
of the vorticity into a quadrupolar structure is closely related to the
spatial organization of kinetic activity and heating 
near the current sheet.
Turning to the more turbulent hybrid run (second row, Fig. \ref{jzwap}),
we see that the proton temperature increase and the proton
temperature anisotropy have a very patchy behavior.  In this case it
is more efficacious to carry out a statistical analysis, as done by
\citet{FranciSW14}.  

\begin{figure}[!hbt]
\begin{center}
\includegraphics[width=\columnwidth]{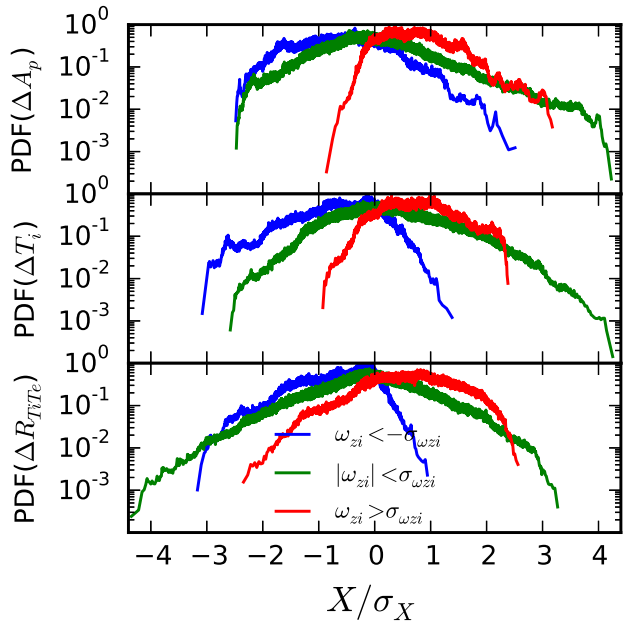}
\caption{Conditional PDFs for the PIC 
run:   (top) $\Delta T_i$
(middle) $\Delta A_p$; and 
(bottom) $T_p/T_e$.
In each case the pdfs are computed for three rages of 
vorticity values: 
(i) $\omega < -\sigma$; 
(ii) $-\sigma < \omega < + \sigma$, and 
(iii) $\omega > \sigma$,
where $\sigma $ is the r.m.s. value of vorticity.
}
\label{pdfspic}
\end{center}
\end{figure}

Therefore, for the turbulent cases, we examine the
association of kinetic effects with $\m{\omega}\cdot\m{B}_0$, by computing
conditional probability distribution functions (PDFs).  For example we
may look at $\Delta A_p$, $\Delta T_i$ and $\Delta R_{TiTe}$, extracting
separate distributions, from the PIC simulation data depending on selected
ranges of the values of $\omega_z$.  These ranges are (i) $\omega <
-\sigma_{\omega zi}$; (ii) $-\sigma_{\omega zi} < \omega < \sigma_{\omega zi}$; and and finally (iii) $\omega
> \sigma_{\omega zi}$ where $\sigma_{\omega zi}$ is the rms value of proton vorticity.  The resulting
distributions for the PIC run Turb812 are shown in Figure \ref{pdfspic} where
one sees immediately that the most probable values of $\Delta A_p$,
$\Delta T_i$ and $\Delta R_{TiTe}$ systematically change with the
selected ranges of vorticity.  Large positive $\omega_z$ gives enhanced
likelihood of hotter protons, more anisotropic protons and protons hotter
than electrons.  For highly negative $\omega_z$ one finds colder protons,
lower anisotropy (parallel dominated) and relatively hotter electrons.
This indicates a clear correlation between higher and anisotropic proton
heating with positive ${\bf \omega}\cdot {\bf B}_0$, consistent with
the findings derived from inspection of the Orszag Tang PIC results
described above, and consistent with the findings of \citet{FranciSW14}
based on hybrid simulation.

\begin{figure}[!hbt]
\begin{center}
\includegraphics[width=\columnwidth]{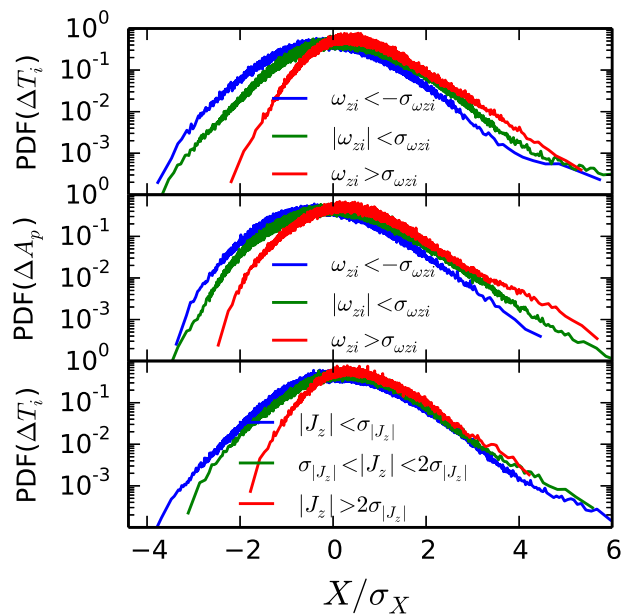}
\caption{Conditional PDFs for the hybrid run. (top) $pdf(T_i)$ conditioned on three ranges of vorticity; 
(middle) $pdf(\Delta A_p$ conditioned on three ranges of vorticity;
(bottom) $pdf(\Delta T_i$ conditioned on three ranges of 
current density. 
}
\label{pdfshybrid}
\end{center}
\end{figure}

The PIC simulation has a relatively complete description of the 
physics but is necessarily of smaller system size (25.6 $d_i$)
and therefore smaller effective Reynolds number. 
For a higher Reynolds number view of the physics in Fig. 
(\ref{pdfspic}), we turn again to a hybrid run 
(204.8 $d_i$).  
Conditional distributions for our hybrid run are shown in
Fig. \ref{pdfshybrid}.  In this case (top to bottom), we plot the
distribution of proton temperatures $\Delta T_i$ conditioned on ranges
of vorticity, the distribution of proton temperature anisotropy
conditions of ranges of vorticity, and the distribution of $\Delta T_i$
conditioned on ranges of the $|J_z|$. 
Here we find that proton heating is associated with both vorticity and
current density but that the association with vorticity is the stronger
of the two effects.

\begin{figure}[!hbt]
\begin{center}
\includegraphics[width=\columnwidth]{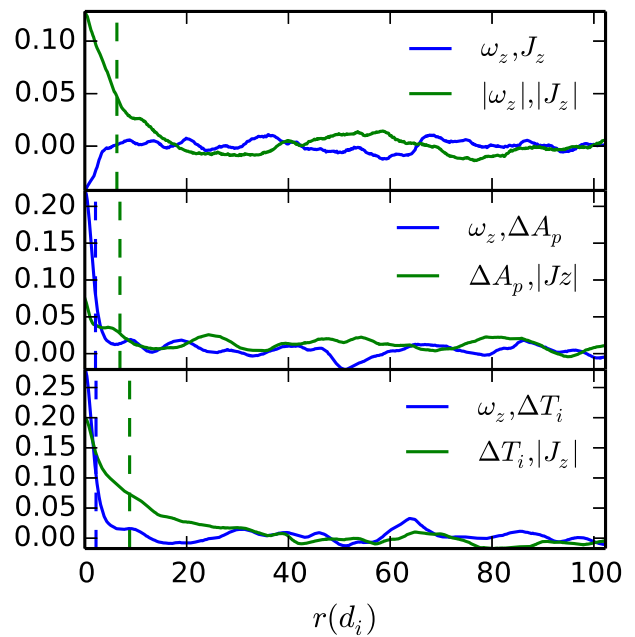}
\caption{
Correlations between multiple quantities for the hybrid run.}
\label{corshybrid}
\end{center}
\end{figure}

\begin{figure}[!hbt]
\begin{center}
\includegraphics[width=\columnwidth]{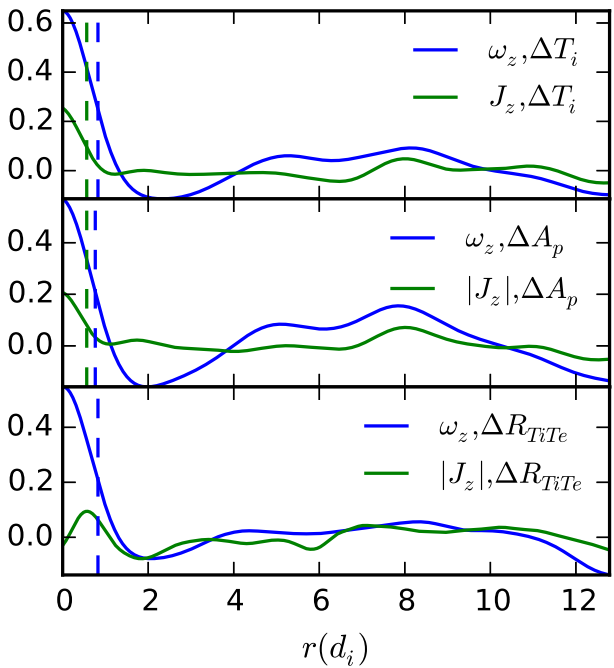}
\caption{
Spatial correlations 
involving vorticity and current density, for PIC runs. (Top) 
proton temperature increase, correlated with current density
and with vorticity, vs, spatial lag; 
(middle) vorticity and current correlations 
with proton temperature anisotropy; 
(bottom) Correlations with 
change in normalized proton to electron temperature ratio.
}
\label{corspic}
\end{center}
\end{figure}

Having confirmed the expected statistical relations between
vorticity and proton kinetic effects for the turbulent cases, 
it remains to quantify the spatial relationships
between current sheets, vorticity concentrations, 
and kinetic effects. We accomplish this using 
two point correlations computed over the whole simulation domain.  
Figure \ref{corshybrid} and Figure \ref{corspic} 
show such correlations for the hybrid and PIC runs, respectively. 

The top panel shows correlations between vorticity and current for the
hybrid run. As current sheets are typically accompanied by vorticity
generated in a quadrupole configuration, we find, on average, that the
correlation between signed $\omega_z$ and $J_z$ is close to zero for
most part.

In contrast, correlation curve for the {\it unsigned} $\omega_z$ and $J_z$
clearly peaks at zero and drops to $1/e$ value in about $6d_i$ as shown by
the vertical green dashed line in Fig. \ref{corshybrid}.  This indicates
the close juxtapostion of vorticity quadrupoles and current sheets.
The numerical resolution in this simulation is $\Delta x = 0.1 d_i = 0.5
d_e$. Because of the coarse grid, the current sheets generally do not
collapse to electron scales and hence the correlation between current
sheets and vorticity goes to $\sim 6 d_i$. The correlations for signed
vorticity and $\Delta T_i$ drops in a few $d_i$s but the correlation of
$\Delta T_i$ with current drops in about $6d_i$ like the correlation
between vorticity and current. This is a direct quantification of the
assertion that vorticity and kinetic effects, such as proton heating,
occur
``near" current sheets and that the vorticity structures are prime
locations of energy exchange between fields and particles. The final panel
of Fig. (\ref{corshybrid})
shows the strong correlation between vorticity and proton anisotropy.

For completeness, we show similar correlations for the PIC run Turb812
in Figure \ref{corspic}. The correlations between vorticity and proton
temperature as well as anisotropy drops to $1/e$ value in less than a
$d_i$. This simulation had extremely fine grid scale, affording the
possibility of collapse of current sheets to electron scales. Hence
the correlations of proton heating and anisotropy to vorticity and
current are much tighter than the hybrid run. The final panel of Figure
\ref{corspic} shows the correlation between current/vorticity and $\Delta
R_{TiTe}$ which also peaks at zero and drops to $1/e$ value in less
than a $d_i$. There is no clear correlation between $\Delta R_{TiTe}$
and current. This again 
is a direct quantification of the fact that protons
are heated ``near" current sheets and not exactly ``in" current sheets
\cite{DmitrukApJ04,ServidioPRL12,MatthaeusApJL16}.

\begin{figure}[!hbt]
\begin{center}
\includegraphics[width=\columnwidth]{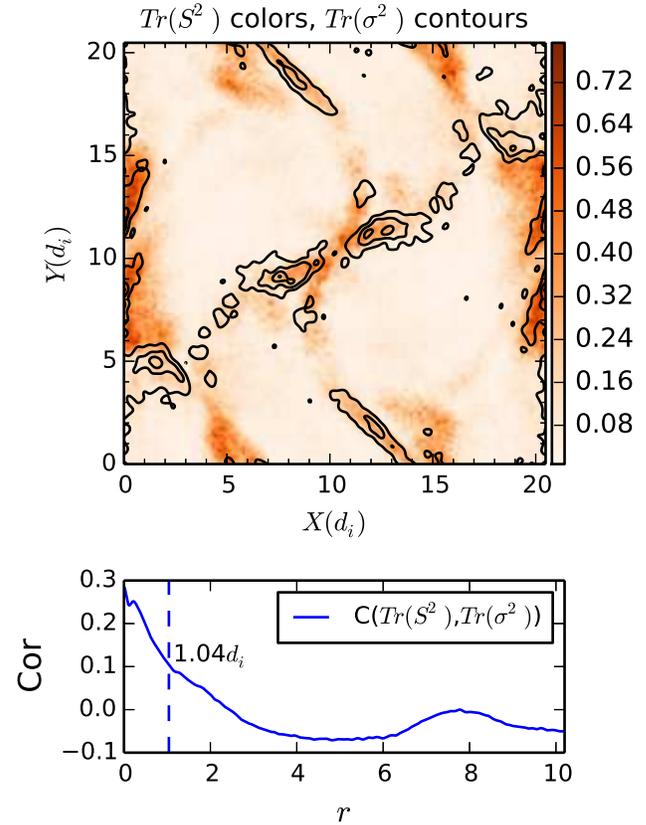}
\caption{
   Second invariants of the symmetric ($S$ in color) and antisymmetric
   ($\sigma$ black contours) stress tensors at initial time evolution of
   the PIC OTV simulation. The two can be seen to have strong proximity
   to each other. The proximity is further quantified by the two point
   correlation of these variables. It drops to $1/e$ value in about $1d_i$
   as shown by the vertical dashed blue line.
}
\label{dsq}
\end{center}
\end{figure}

\section{ Discussion }
In this paper we have examined several ways in which vorticity enters
into the dynamical activity near current sheets, thus contributing
to an increasingly familiar scenario in which sharp gradients and
coherent structures contribute to dissipation, kinetic effects including
anisotropies, and heating.

The most studied cases of coherent structures in space
plasmas have been current sheets or reconnection sites
\citep{SorrisoValvoGRL99,GrecoGRL08,OsmanApJL14}, at least in part
due to the availability of magnetic field data from spacecraft with
sufficient resolution to attain accurate detection at scales at or
approaching kinetic scales.  One would fully expect however that
additional plasma variables such as velocity (and density, 
not shown here; however see \citet{ServidioApJL14})
will be
involved in intermittency and formation of coherent structures.
In fact there have been numerous recent indications that velocity
gradients might be responsible for heating the plasma through a
linear instability mechanism \citep{MarkovskiiApJ06} or a viscous
like interaction \citep{VasquezApJ12}.  The conclusion that velocity
gradients contribute to proton kinetic effects is supported by recent
simulation \citet{FranciSW14}, and based on solar wind observation
\citep{ServidioApJL14}. Similar conclusions may be extended to sharp
gradients in density as well as velocity and magnetic field.

In this context we may summarize our main conclusions, and suggest a
broader context.  Kinetic plasma turbulence forms current sheets, which like
their MHD counterparts, generate vorticity nearby, in regions of strong
current gradients.  Various kinetic effects and heating of both protons
and electrons take place in the vicinity of these current and vortex
structures. We have shown here (see also \citet{FranciSW14})
that proton heating and anisotropy is better associated with vorticity rather
than current, which provides some clarification to results that have
shown \citep{ServidioPRL12,GrecoPRE12} that various kinetic
effects are concentrated in the general vicinity of current sheets.

The idea that kinetic effects concentrate near current
sheets in turbulence might easily extend
to include effects such as flux
pile-up, density compressions and particle trapping in secondary islands. 
All of these 
are likely features of the dynamics when current sheets
form between interacting magnetic flux tubes.  In this way, it may be
useful to think of the the general region within a few $d_i$ neighborhood
a strong current sheet as a kind of complex or generalized coherent
structure involving juxtaposition of coherent
structures in several variables -- current, density, vorticity,
compressions, heating to name a few.  This agglomeration 
has been demonstrated to span a range of scales, 
e.g., from proton to
electron inertial scales in the shear driven turbulent current sheets
reported by \citet{KarimabadiPP13}.

For large systems sizes and large effective Reynolds numbers it is likely
that the dynamical interactions among several types of coherent structures
become an important feature of the overall dynamics.  In that case a
fuller understanding of the complex dynamics within a ``generalized
coherent structure'' becomes essential.  Here we described a first step -- 
a simple dynamical chain that leads from current sheet formation to vorticity
quadrupole generation and emergence of enhanced proton kinetic effects.

One can actually see readily that even this situation must be more complex
than we have described so far: From Eq. \ref{eq:random} we can deduce
(see also \cite{DelSartoPRE16})
that increase of internal energy, or heating in the collisonless limit
must be associated with the full contraction of the pressure tensor
with the gradient tensor $\nabla_iu_j$.  However, the pressure tensor
is symmetric and gradient tensor may be written as as sum of symmetric
strain and antisymmetric strain, the latter containing the vorticity.
Therefore the only the symmetric stress and not the vorticity  
contributes directly to heating in the absence of
collisions.  So why do we see an association of vorticity and heating?
For an empirical answer we refer to Fig \ref{dsq} which shows the
second invariant of the symmetric stress ($Det(S^2)$) and a measure
of vorticity $Tr(\sigma^2)$
plotted in the simulation plane for the PIC OTV case at
the same time as in Figure \ref{jzwap}. Evidently there is a very high
degree of correlation.  The symmetric stress can produce heating, while
vorticity can produce thermal anisotropy \cite{DelSartoPRE16,HubaGRL96}.
Mathematically the two effects can be co-located even if they are formally
independent because the current sheets are locally quasi-one-dimensional
objects.

   In this paper we have discussed the spatio-temporal relationship
   between current sheets, vorticity and kinetic processes such as 
   proton heating, electron heating and anisotropy. 
   As mentioned above, 
   other quantities (e.g. density, compressions, localized
   linear waves, shock(lets), pressure dilations) also may be 
   intermittent and
   may have strong correlations with current sheets. A detailed study of
   such processes is beyond the scope of this paper and will be discussed
   elsewhere. A detailed mathematical study of the relationship between
   symmetric and antisymmetric parts of the stress tensor is also beyond
   the scope of this study. An excellent start in that direction is the
   paper by \citet{DelSartoPRE16}. Further study 
 will be needed to tighten the simple picture presented in this paper.

\begin{acknowledgments}
Research is supported by NSF AGS-1063439, AGS-1156094 (SHINE), AGS-1460130
(SHINE),  and NASA grant NNX14AI63G (Heliophysics Grandchallenge Theory),
the MMS mission through grant NNX14AC39G, and the Solar Probe Plus science
team (ISIS/SWRI subcontract No. D99031L).  Some of the simulations were
performed at the NCAR Yellowstone supercomputer.
\end{acknowledgments}


\end{document}